\documentclass[pre,twocolumn,showpacs,superscriptaddress,preprintnumbers,amsmath,amssymb]{revtex4-1}
\usepackage{footnote}
\usepackage{graphicx}
\usepackage{dcolumn}
\usepackage{bm}
\usepackage{color}
\usepackage{xcolor}
\usepackage{ulem}
\usepackage{courier}
\usepackage{braket}
\usepackage[T1]{fontenc}
\usepackage[ansinew]{inputenc}
\usepackage{longtable}

\begin{document}

\title{Carbon ionization at Gbar pressures: an \textit{ab initio} perspective on astrophysical high-density plasmas}

\author{Mandy Bethkenhagen}
\affiliation{Institut f\"ur Physik, Universit\"at Rostock, 18051 Rostock, Germany}
\affiliation{Laboratoire de G\'{e}ologie de Lyon, \'{E}cole Normale Sup\'{e}rieure de Lyon, 69364 Lyon Cedex 07, France}
\author{Bastian~B.~L.~Witte}
\affiliation{Institut f\"ur Physik, Universit\"at Rostock, 18051 Rostock, Germany}
\affiliation{SLAC National Accelerator Laboratory, 2575 Sand Hill Road, MS 72 Menlo Park, CA 94025 USA}
\author{Maximilian~Sch\"orner}
\affiliation{Institut f\"ur Physik, Universit\"at Rostock, 18051 Rostock, Germany}
\affiliation{SLAC National Accelerator Laboratory, 2575 Sand Hill Road, MS 72 Menlo Park, CA 94025 USA}
\author{Gerd~R\"opke}
\affiliation{Institut f\"ur Physik, Universit\"at Rostock, 18051 Rostock, Germany}
\author{Tilo~D\"oppner}
\affiliation{Lawrence Livermore National Laboratory, Livermore, CA 94550, USA}
\author{Dominik~Kraus}
\affiliation{Helmholtz-Zentrum Dresden-Rossendorf, 01328 Dresden, Germany}
\affiliation{Institute of Solid State and Materials Physics, Technische Universit\"at Dresden, 01069 Dresden, Germany}
\author{Siegfried~H.~Glenzer}
\affiliation{SLAC National Accelerator Laboratory, 2575 Sand Hill Road, MS 72 Menlo Park, CA 94025 USA}
\author{Philip~A.~Sterne}
\affiliation{Lawrence Livermore National Laboratory, Livermore, CA 94550, USA}
\author{Ronald~Redmer}
\affiliation{Institut f\"ur Physik, Universit\"at Rostock, 18051 Rostock, Germany}

\begin{abstract}
A realistic description of partially-ionized matter in extreme thermodynamic states is critical to model the interior and evolution of the multiplicity of high-density 
astrophysical objects. Current predictions of its essential property, the ionization degree, rely widely on analytical approximations that have been challenged recently 
by a series of experiments. Here, we propose a novel \textit{ab initio} approach to calculate the ionization degree directly from the dynamic electrical conductivity 
using the Thomas-Reiche-Kuhn sum rule. This Density Functional Theory framework captures genuinely the condensed matter nature and quantum effects typical for 
strongly-correlated plasmas. We demonstrate this new capability for carbon and hydrocarbon, which most notably serve as ablator materials in inertial confinement fusion 
experiments aiming at recreating stellar conditions. We find a significantly higher carbon ionization degree than predicted by commonly used models, yet validating the 
qualitative behavior of the average atom model Purgatorio. Additionally, we find the carbon ionization state to remain unchanged in the environment of fully-ionized hydrogen.
Our results will not only serve as benchmark for traditional models, but more importantly provide an experimentally accessible quantity in the form of the electrical 
conductivity.
\end{abstract}

\maketitle

\section{Introduction}
\label{sec:introduction}

Modeling the internal structure and thermal evolution of low-mass stars, brown dwarfs, and massive giant planets 
requires accurate equation of state data and even more importantly reliable transport properties of warm dense 
matter~\cite{Chabrier1997, Becker2018}. For example, the interplay of convective and radiative transport in low-mass stars
is reflected by key plasma quantities such as opacity, electrical conductivity, and absorption coefficients. All those properties 
can be directly linked to the ionization degree, which is defined as the ratio between the number of free electrons and the sum of all electrons.

The ionization degree can be obtained directly from the Saha equations for the limiting case of the low-density plasma in thermodynamic equilibrium. 
In this framework, the corresponding ionization energies are defined as the difference between the ground state energy and its continuum of free states.
Generally, the ionization energies crucially depend on the temperature and density of a plasma. For example, an increase of the density results in a lowering of the 
ionization energies with respect to their well-known values for isolated atoms due to correlation effects such as screening of the Coulomb interaction, self-energy, 
strong ion-ion interactions, and Pauli blocking~\cite{ZKKKR, KKER}. This effect is known as Ionization Potential Depression (IPD) and inherit to any theory aiming
at predicting the ionization degree, which has been subject of many-particle physics for decades~\cite{KKER, OPAL, Murillo1998, Crowley2014}. 
For instance, the simple Debye-H\"uckel theory for static screening has been combined with the ion sphere model by Ecker and Kr\"oll (EK)~\cite{Ecker1963} and later 
improved by Stewart and Pyatt (SP)~\cite{Stewart1966}. The predictions of both models differ considerably for high-density plasmas as encountered in the deep interior 
of astrophysical objects, which are characterized by pressures up to the Gbar range and temperatures of several eV to keV. 

Matter under such extreme conditions is notoriously challenging to produce and probe. However, great advances in X-ray techniques have been made over the last decade and
have been implemented at high-power laser facilities and free electron lasers (FELs), which are now available for the experimental study of high-density plasmas.
For instance, the ionization state of isochorically heated solid aluminum was extracted at temperatures in the range of 10-100~eV by measuring the $K$ edge threshold
at the Linac Coherent Light Source (LCLS)~\cite{Vinko2012, Ciricosta2012, Cho2012}. Additionally, the ionization of hot dense aluminum plasmas in the range of 
1-10~g/cm$^3$ and 500-700~eV was determined using the ORION laser~\cite{Hoarty2013}. Experiments performed at the Omega laser facility and National Ignition Facility 
compressed hydrocarbon (CH) up to 100~Mbar, and obtained the ionization state via X-ray Thomson scattering (XRTS)~\cite{Fletcher2014, Kraus2016}. The same technique was applied at the 
LCLS to measure the IPD in carbon plasma~\cite{Kraus2018}. Furthermore, the total intensity of plasma emission in Al and Fe driven by narrow-bandwidth X-ray pulses across a 
range of wavelengths was utilized to determine the IPD~\cite{Kasim2018}. Generally, the results of those experiments indicate that rather simple models including IPD based 
on the EK or SP approaches fail to describe the ionization degree correctly~\cite{Crowley2014}. 

Therefore, novel theoretical concepts for the prediction of the ionization degree are imperatively required and first improvements have been made. For example, a 
two-step Hartree-Fock-Slater (HFS) approach has been proposed recently~\cite{Son2014}. It is a combined atomic-solid-plasma model that permits ionization potential 
depression studies also for single and multiple core hole states~\cite{Rosmej2018}, or the dynamic ion-ion structure factor~\cite{Lin2017}. The latter approach has 
been generalized to include Pauli blocking effects which are important at high densities~\cite{Roepke2019}.
Another route is to characterize ionization by applying molecular dynamics simulations for the ions in combination with electronic structure calculations using density 
functional theory (DFT-MD)~\cite{Vinko2014, Hu2017, Driver2018}, which is especially well-suited for dense plasmas. So far, all DFT-MD works relied widely on the density of 
states (DOS), which was used to analyze the evolution of the ionization degree with density and temperature~\cite{Vinko2014, Hu2017, Driver2018}. However, none of these 
works provided a consistent picture of the ionization degree resolving the recent discussion on IPD in high-density plasmas; see Refs.~\cite{Iglesias2014, Hu2017, Roepke2019}, 
the comment of Iglesias and Sterne~\cite{Iglesias2018}, and the reply by Hu~\cite{Hu2018}. This debate is fundamentally related to the question of how to define the 
ionization degree properly for warm dense matter, which is characterized by densities beyond the applicability range of the Saha equations and its underlying chemical 
picture.

In this work, we meet this challenge by calculating the ionization degree directly from an experimentally accessible quantity: the dynamic electrical conductivity. This
DFT-MD method takes into account the electronic and ionic correlations in a self-consistent way and, in particular, reflects essential features of high-density 
plasmas such as the existence of energy bands instead of sharp atomic levels and their occupation according to Fermi-Dirac statistics. In contrast to a definition relying
entirely on the DOS, our novel approach is based on the Thomas-Reiche-Kuhn (TRK) sum rule and the evaluation of electronic transitions originating solely from electrons within 
the conduction band.
This method, which has, to our knowledge, never been used before, is a step toward precisely modeling matter at high energy densities as occurring, 
e.g., in inertially confined fusion experiments~\cite{Glenzer2012short} or in low-mass stars~\cite{Chabrier1997}, brown dwarfs, and massive giant planets~\cite{Becker2018}. 
Carbon and CH are chosen as exemplary materials relevant to the before-mentioned applications.

\section{Methods}
\label{sec:methods}
\subsection{Deriving ionization from the dynamic conductivity and the sum rule}
\label{subsec:theory-conductivity}
The dynamic electrical conductivity, also referred to as optical conductivity, is calculated from the Kubo-Greenwood formula, 
\begin{align}
\begin{split}
 \sigma^{\textrm{tot}} (\omega)\ &=\frac{2\pi e^{2}}{3V\omega}\sum_{\mathbf{k}\nu\mu} \left(f_{\mathbf{k}\nu} - f_{\mathbf{k}\mu}\right) \left| \bra{\mathbf{k}\nu}\hat{\mathbf{v}}\ket{\mathbf{k}\mu}\right|^2 \cr &  \times\delta\left(E_{\mathbf{k}\mu}-E_{\mathbf{k}\nu}-\hbar\omega\right);
\end{split}
\label{eq:kubo-greenwood}
\end{align}
which can be derived from linear response theory~\cite{Kubo1957, Greenwood1958, Recoules2005, Holst2011}. 
In the above equation, the transition matrix elements $\left|\bra{\mathbf{k}\nu}\hat{\mathbf{v}}\ket{\mathbf{k}\mu}\right|^2$ with the velocity operator $\hat{\mathbf{v}}$ are the key components. They reflect the transition probability between the initial eigenstate associated with band $\nu$ and the final eigenstate in band $\mu$ at a particular \textbf{k} point in the Brillouin zone of the simulation box of volume $V$. For a given frequency $\omega$, only states with a positive difference between eigenenergies $E_{\mathbf{k}\mu}$ and $E_{\mathbf{k}\nu}$ contribute to the conductivity. The occupation of initial and final states is weighted with the Fermi-Dirac function 
$f_{\mathbf{k}\nu} = \left[ \exp((E_{\mathbf{k}\nu}-\mu_{\textrm{e}})/k_B T)+ 1 \right]^{-1}$, whereas $T$ and $\mu_{\textrm{e}}$ denote the temperature and the chemical potential of the electrons, respectively. Additionally, $e$ and $\hbar$ represent the elementary charge and the reduced Planck constant in Eq.~\eqref{eq:kubo-greenwood}. 

The resulting dynamical electrical conductivity has to fulfill the well-known Thomas-Reiche-Kuhn (TRK) sum 
rule for dipole transitions~\cite{Thomas1925, Reiche1925, Kuhn1925, Mahan, Desjarlais2002},
\begin{equation}
 Z^{\textrm{tot}} = \frac{N^\textrm{tot}_\textrm{e}}{N_{\textrm{i}}} =\frac{2m_e V}{\pi e^2 N_{\textrm{i}}} \int_{0}^{\infty} d\omega \;\sigma^{\textrm{tot}}(\omega)\;.
\label{eq:sumrule}
\end{equation}
It yields the ratio between the total number of electrons $N^\textrm{tot}_\textrm{e}$ and the number of ionic centers in the system 
$N_{\textrm{i}}$ and establishes an important convergence criterion for the numerical computation of the dynamic electrical conductivity.
For the examples chosen in this work, we require charge state values of $Z^{\textrm{tot}}=6$ for carbon and $Z^{\textrm{tot}}=7$ for CH in order to fulfill the TRK sum rule 
exactly. The number of ionic centers is in both cases $N_\textrm{i} = N_\textrm{C} = N_\textrm{CH} = 32$. 

Our novel approach separates the dynamic electrical conductivity $\sigma^{\textrm{tot}}$ into three individual parts based on the different nature of electronic transitions 
in the energy spectrum. Hence, the transition matrix elements for a given $\textbf{k}$ point of the sum in Eq.~\eqref{eq:kubo-greenwood} are divided into contributions 
attributed to intraband transitions in the conduction (c-c) and valence (v-v) band as well as interband transitions between the valence and conduction band (v-c):
\begin{equation}
 \sigma^{\textrm{tot}}(\omega) = \sigma^{\textrm{c-c}}(\omega) + \sigma^{\textrm{v-c}}(\omega) + \sigma^{\textrm{v-v}}(\omega)\;,
\label{eq:sigma-parts}
\end{equation}
whereas each contributions $\textrm{x}=\{\textrm{c-c, v-c, v-v}\}$ is required to fulfill the partial TRK sum rule
\begin{equation}
 Z^{\textrm{x}} \equiv \frac{2m_e V}{\pi e^2 N_{\textrm{i}}} \int_{0}^{\infty} d\omega\; \sigma^{\textrm{x}}(\omega).
 \label{eq:Zx}
\end{equation}
The individual conductivity contributions can be simply identified by choosing an energy within the energy gap between the valence and conduction band, which is chosen 
naturally as the center of the gap between the states of interest. For our carbon and CH examples, the gap is always chosen between the clearly identifiable 1s valence and 
the continuum as conduction states, which already comprise the 2s and 2p states at the considered conditions. 

The electrons effectively contributing to the conductivity in the conduction band are the free electrons $N_e^{\textrm{free}}$, so that we can identify 
$\sigma^{\textrm{x}}(\omega) = \sigma^{\textrm{c-c}}(\omega)$ in Eq.~\eqref{eq:Zx} and define the ionization state as
\begin{equation}
 Z^{\textrm{free}} = \frac{N^{\textrm{free}}_{\textrm{e}}}{N_{\textrm{i}}} \equiv Z^{\textrm{c-c}}, 
 \label{eq:Zfree}
\end{equation}
which we propose as a new and suitable measure for this quantity in high-density plasmas. 

Finally, we calculate the ionization degree, 
\begin{equation}
 \alpha = \frac{Z^\textrm{free}}{Z^\textrm{tot}} = \frac{N^\textrm{free}_\textrm{e}}{N^\textrm{tot}_\textrm{e}},
  \label{eq:alpha}
\end{equation}
which is consequently defined as the ratio between the number of free and total charge carriers per number of ionic centers $N_\textrm{i}$.

\subsection{Computational details}
\label{subsec:computational}
The DFT-MD simulations for carbon and CH were performed with the 
Vienna Ab initio Simulation Package (VASP)~\cite{Kresse1993, Kresse1994, Kresse1996}.
We considered densities between 20-400~g/cm$^3$ at a temperature of 100~eV resulting in a 
pressure range between 0.8-65~Gbar. These conditions correspond to a maximum 
compression ratio of more than 100 and thus, it was crucial to treat all electrons 
explicitly using the Coulomb potential with a cutoff energy of 15~keV. We considered 32~carbon atoms up to 150~g/cm$^3$
and 64~carbon atoms for the three highest densities starting at 200~g/cm$^3$. For the CH calculations, we added 
32 and 64 hydrogen atoms to the respective pure carbon simulations. Additionally, a large 
number of bands, i.e. typically 800 -- 5000 bands, was required to describe the adequately at the high temperature considered here. Each DFT-MD 
simulation was run for at least 20~000 time steps with a step size of 50~as for carbon and 10~as for
CH in order to reflect the ion dynamics properly. A Nos\'{e}-Hoover thermostat~\cite{Nose1984} was used to control the 
ion temperature, the Brillouin zone was evaluated at the Baldereschi mean value point~\cite{Baldereschi1973},
and we employed the exchange-correlation (XC) functional of Perdew, Burke, and Ernzerhof (PBE)~\cite{Perdew1996}.\\  
The electrical conductivity was determined from an average over 20~snapshots taken from the DFT-MD simulation per condition, and a Monkhorst-Pack 2x2x2 
grid was used to evaluate the Kubo-Greenwood formula, Eq.~(\ref{eq:kubo-greenwood}).
We performed extensive convergence tests of the DFT-MD simulations with respect to the energy cutoff, number of bands, \textbf{k} point sets, and the number 
of atoms. Furthermore, we tested the influence of the XC functional used in the DFT cycles by carrying out
additional calculations with the LDA and SCAN~\cite{Sun2015} functionals. All parameters were chosen such
that the TRK sum rule is always fulfilled within 2\%, which depends most significantly on the number of explicitly considered bands. 


\section{Results}
\label{}

\subsection{Dynamic electrical conductivity and sum rule}
\label{sec:dynsigma}
\begin{figure}[b!]\includegraphics[clip=true,width=1.0\linewidth]{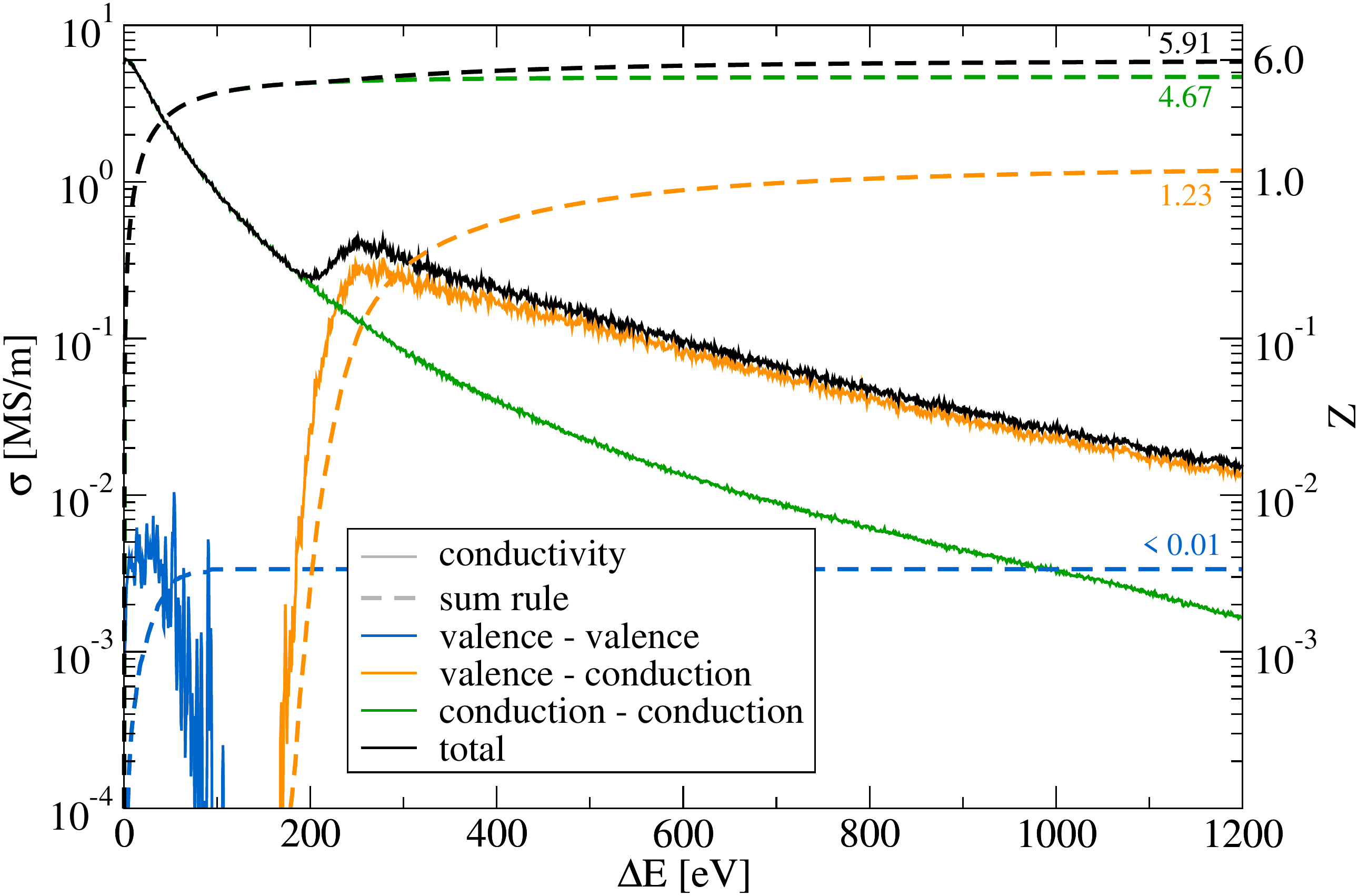} 
\caption{Electrical conductivity (solid lines) and TRK sum rule values (dashed lines) for a carbon simulation snapshot at 50~g/cm$^3$ and $T=100$~eV. The different colors 
indicate the total value and the individual components of both quantities according to Eq.~\eqref{eq:sigma-parts} and Eq.~\eqref{eq:Zx}. The final TRK sum rule values are 
given as colored numbers.
} 
\label{fig:sigma}
\end{figure}

In the following, we are demonstrating our novel method for an exemplary single snapshot of a carbon plasma at 50~g/cm$^3$ and 100~eV. 
In Fig.~\ref{fig:sigma}, the total dynamic electrical conductivity obtained with the Kubo-Greenwood approach is shown
as solid black line. The curve spans three orders of magnitude over the entire considered energy range and exhibits a pronounced local maximum at
about 250~eV, which results from the strong v-c conductivity contribution as becomes apparent upon breaking up the total conductivity into its individual contributions
associated with c-c, v-c, and v-v transitions according to Eq.~(\ref{eq:sigma-parts}). While the c-c contribution dominates the total conductivity at energies below 225~eV, 
the v-c contribution shows a pronounced threshold behaviour at about 175~eV and starts to prevail at energies above 225~eV. At the same time, the v-v transition contribution
is almost negligible and can be associated with hopping processes. These can occur as a result of the partial filling of the 1s states at the extreme densities and temperatures investigated 
here. This behavior is contrary to the known 0~K concept for solids, where the v-v conductivity has to be zero, since the full occupation of the 1s states leads to a vanishing transition 
probability due to selection rules.

Evaluation of the TRK sum rule for the total conductivity according to Eq.~(\ref{eq:sumrule}) yields a value of 5.91, which agrees with the exact sum rule value of 6 within 
2~\%. The same procedure applied to the v-c contribution results in a value of 1.23, which makes about 21~\%  to the total sum rule at these conditions. Additionally, the 
sum rule for the v-v transitions yields a value less than 0.01, which translates to about 0.2~\% of the total value, and contributes the most at low energies. Finally,the 
largest total sum rule contribution of the remaining 79~\% is attributed to the sum rule value applied to the c-c conductivity contribution. The resulting c-c value is 
4.67 and will be identified as the ionization state later in this work. Note, that all sum rule values given in the following sections are corrected by a factor 
$Z^\textrm{exact}/Z^\textrm{tot}$ that accounts for the numerical uncertainty.

\subsection{Density of states}
\begin{figure}[b!]
\includegraphics[width=0.95\linewidth]{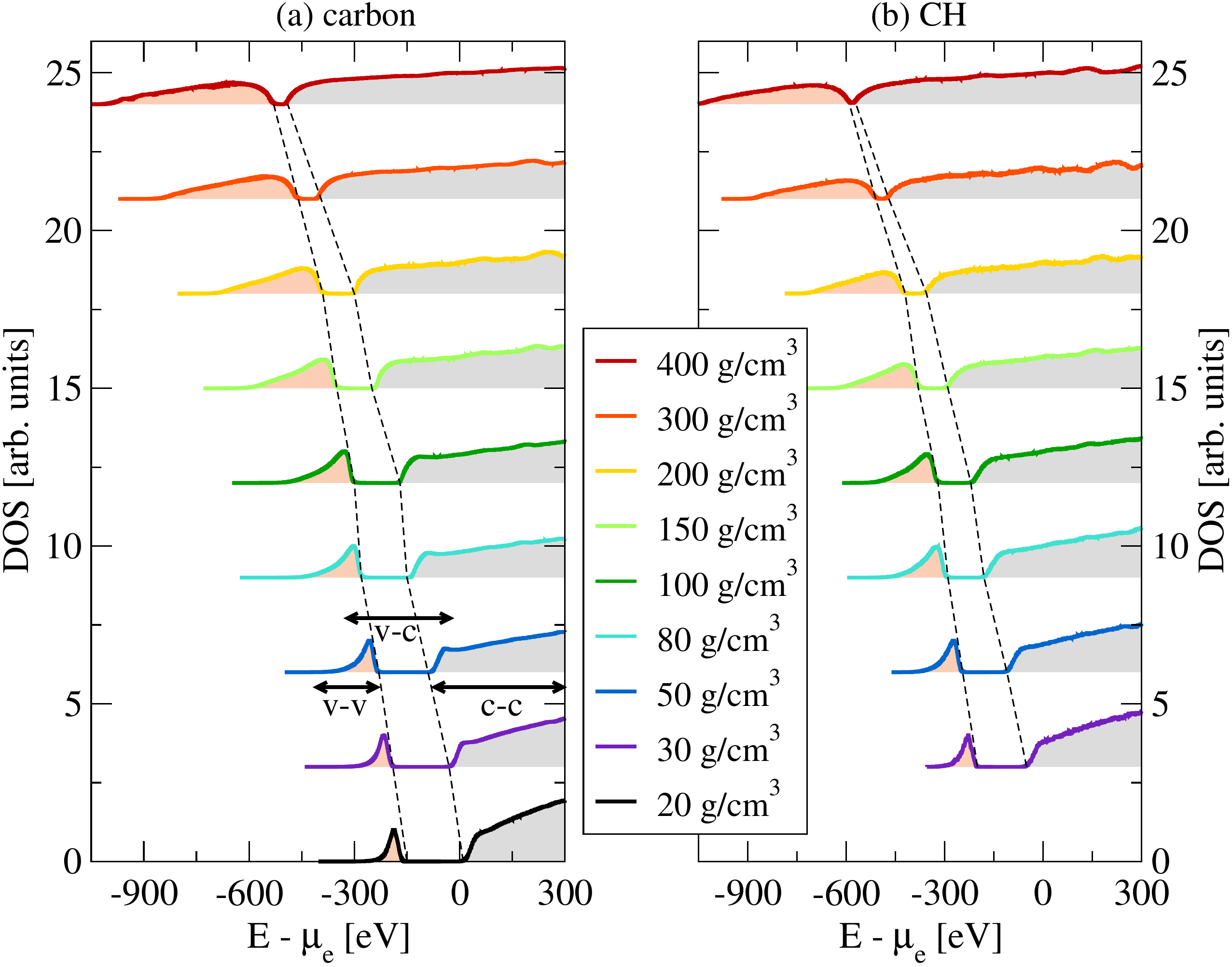} 
\caption{
Density of states of carbon (a) and CH (b) for different densities at T = 100 eV. The pink areas indicate states in the 1s valence band, while states in 
the conduction band are colored in grey. The black arrows show exemplarily at 50 g/cm$^3$ the regions associated with intraband (v-v, c-c) and interband transitions (v-c). 
The dashed lines serve as guide to the eye for the valence-conduction gap.} 
\label{fig:DOS}
\end{figure}

In Fig.~\ref{fig:DOS}, we show our results for the density of states (DOS) for carbon and CH for all considered densities  at 100~eV. Each DOS 
shows a pronounced valence band corresponding to the 1s states at small energies and the continuum of conduction states at high energies. Note that all energies
are plotted with respect to the chemical potential $\mu_e$. 

For both materials, we observe the valence bands to broaden and shift towards smaller energies with increasing
density. At the same time, the edge of the conduction states moves as well towards smaller energies and the gap between valence and conduction bands narrows with
rising density. However, the gap never completely vanishes for the considered conditions and can be still clearly identified at the highest considered density of 
400~g/cm$^3$.

The center of the gap between valence and conduction bands serves as input for our method to calculate the different conductivity contributions. In principle, 
any energy in the gap can be used to separate valence from conduction states. In this work, the gap was determined for every snapshot individually via the smallest energy 
difference between a state in the valence band and one in the conduction state. This approach is formally equivalent to the HOMO-LUMO method, which 
is applied at T=0~K to obtain the energy difference between highest occupied and lowest unoccupied molecular orbital.

\subsection{Conductivity-based Ionization}
\label{subsec:ionization}
\begin{figure}[b!]\includegraphics[width=1.0\linewidth]{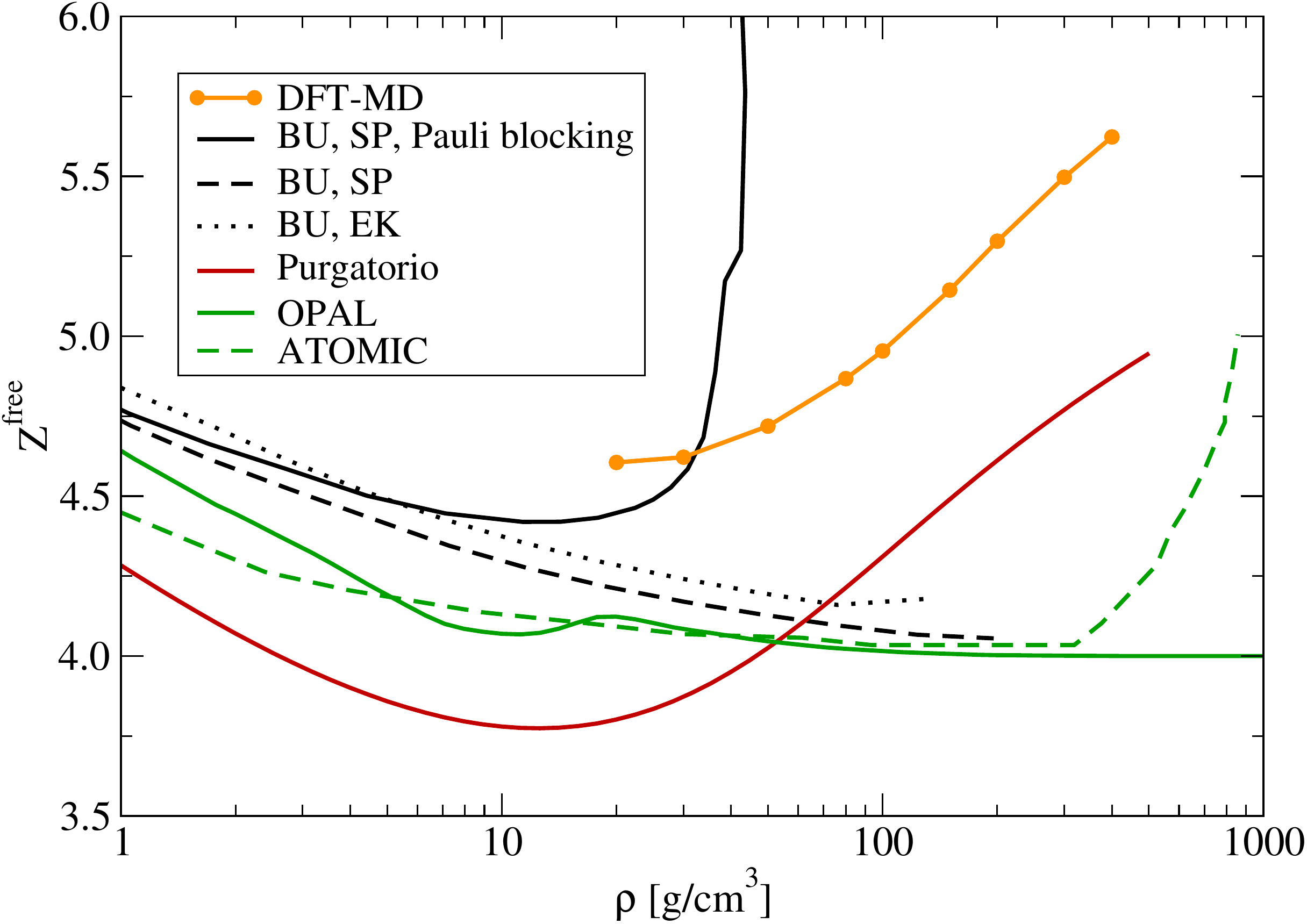} 
\caption{Ionization state of carbon $Z^\textrm{free}$ derived from DFT-MD simulations (orange line) according to Eq.~(\ref{eq:Zfree}) compared to predictions of 
Purgatorio (red line), OPAL (green solid line)~\cite{OPAL}, ATOMIC (green dashed line)~\cite{Hakel2006} and Beth-Uhlenbeck (BU) calculations (black lines)~\cite{Roepke2019}. 
The BU results incorporate the two different IPD models by Ecker-Kr\"oll (EK) and Stewart-Pyatt (SP), respectively, and the solid line additionally takes into account Pauli 
blocking effects.}
\label{fig:ionization-Z}
\end{figure}

In Fig.~\ref{fig:ionization-Z} we present our DFT-MD results for the ionization state of dense carbon for a temperature of 100~eV as function of density compared 
to ionization models that are commonly used for modeling astrophysical or ICF capsule implosions. 

For the lowest densities considered here, the predictions for the carbon ionization state of Purgatorio~\cite{Sterne2007}, ATOMIC~\cite{Hakel2006}, OPAL~\cite{OPAL} and 
the different Beth-Uhlenbeck (BU) models~\cite{Roepke2019} agree qualitatively and predict a decreasing ionization state with increasing density. However, two classes of
models can be identified for the high-density regime. On one hand, OPAL and both BU models without Pauli blocking generally continue that trend at densities 
above 10~g/cm$^3$. Among the three models, OPAL predicts the smallest ionization state and converges towards a constant value of 4. The BU curve including IPD based on
the SP model gives slightly higher ionization states, but leads to no essential change in the general behavior compared to the BU approach using the EK description instead.
On the other hand, the second class of curves, namely Purgatorio, ATOMIC, the BU model including Pauli blocking as well as our DFT-MD results, is characterized by a steep 
rise at high densities, which is associated with pressure ionization expected under those conditions~\cite{Mott61}. However, the slope and onset of this rise in ionization
state vary vastly depending on the approach. The steepest slopes are predicted by ATOMIC and the BU model including Pauli blocking, which in contrast to the other two BU 
models includes electron degeneracy. While ATOMIC suggests the onset of the increase at about 300~g/cm$^3$, the BU model including Pauli blocking predicts this effect at
a density an order of magnitude lower. The DFT-MD results confirm the increase, but the slope of our DFT-MD curve is not as steep. Furthermore, our calculations capture the 
slope of the average atom model Purgatorio, which evaluates the effective charge at the Wigner Seitz radius~\cite{Sterne2007}. However, our \textit{ab initio} calculations 
yield systematically about 0.5 higher ionization states than Purgatorio indicating that this effective one-particle model is not capturing all important correlation effects
treated via our many-body method.

\begin{figure}[htb]\includegraphics[width=1.0\linewidth]{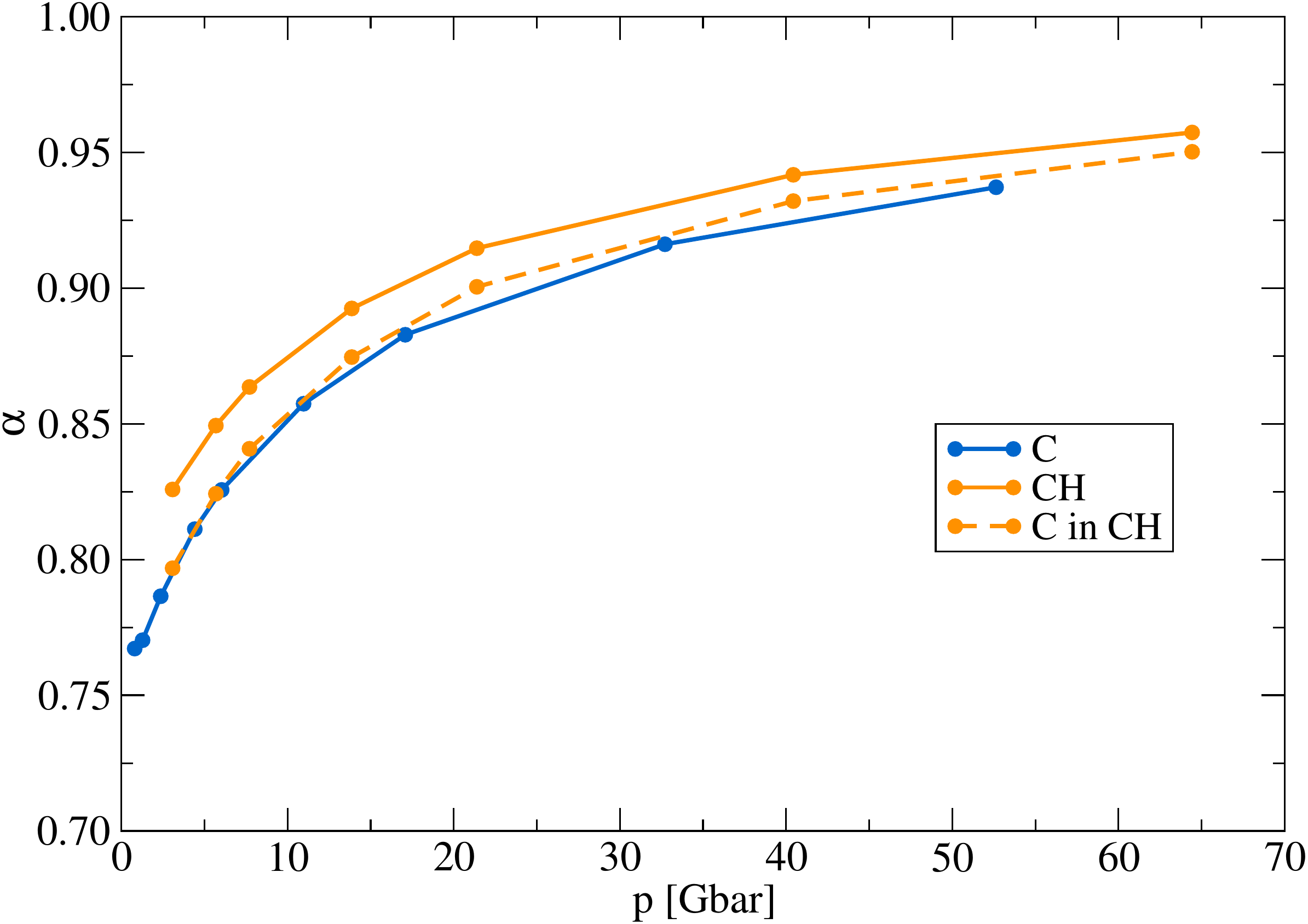} 
\caption{Ionization degree $\alpha$ of pure carbon (blue line), CH (solid orange line), and carbon in CH assuming hydrogen to be fully-ionized (dashed orange line).}
\label{fig:ionization-alpha}
\end{figure}

The ionization degree derived according to Eq.~\eqref{eq:alpha} is shown as a function of pressure for carbon and CH  
as solid lines in Fig.~\ref{fig:ionization-alpha}. For both materials, we find a steady increase of the ionization degree in the range of 0.76 to 0.96. Both curves have a 
very similar slope, while CH yields slightly higher ionization degrees than carbon. Additionally, we plot the ionization degree of the carbon in CH as dashed curve, which 
agrees remarkably well with the values found for pure carbon indicating that the ionization degree of carbon is not changed by adding hydrogen. Note, that this curve was 
calculated under the assumption that all hydrogen atoms are fully ionized. This assumption was tested for pure hydrogen at 80~g/cm$^3$, where we find an ionization degree 
of 1.00 as a result of the conductivity showing only a c-c contribution and of the vanished gap in the DOS.

\section{Conclusion}
\label{sec:conclusion}
Our presented novel method to derive the ionization state and ionization degree entirely from \textit{ab initio} simulations by applying the exact TRK sum 
rule for the dynamic electrical conductivity is entirely self-consistent. This approach naturally takes into account the electronic and ionic correlations, 
in particular, it reflects essential features of high-density plasmas such as the existence of energy bands instead of sharp atomic levels and their occupation 
according to Fermi-Dirac statistics. Our DFT-MD results for the carbon ionization state predict a gradually increasing pressure ionization and strongly disagree 
with commonly used models such as BU~\cite{Roepke2019}, OPAL~\cite{OPAL}, and ATOMIC~\cite{Hakel2006} in high-density plasmas, yet, we confirm the slope of the average 
atom model Purgatorio~\cite{Sterne2007}. This indicates that the degeneracy and many-body effects contained in our DFT-MD treatment are crucial to incorporate in 
any description of the ionization degree and that assumptions based on atomic physics are not valid to treat the ionization balance in high-density plasmas properly. 
Our conductivity-based method directly exploits knowledge of possible electronic transitions taking into account the nature of the wavefunctions of different
states, which cannot be captured by a method that solely relies on the evaluation of the density of states, and additionally provides an experimentally accessible
quantity. Hence, the presented data will be useful for analyzing and predicting conditions in inertial confinement fusion experiments using, e.g., the National 
Ignition Facility. In particular, this data will guide the understanding of XRTS spectra; see~\cite{Witte2017Al, Kraus2016b}. 

Finally, the conditions considered in this work, are typically found in high-density astrophysical objects. For instance, densities of 100~g/cm$^3$ and temperatures of 
100~eV are expected in the interior of M~dwarf stars with a mass of 0.1~M$_\odot$ (in units of the solar mass). Our results for the ionization degree of carbon and CH 
can be directly used as input for interior structure models, whose underlying radiation transport models and the nuclear reaction rates crucially rely on ionization 
models and opacities.



\section*{Acknowledgements}
We thank Luke B. Fletcher, Martin French, Dirk O. Gericke, Clemens Kellermann, Laurent Masse, and Martin Preising for helpful discussions. M.B., B.B.L.W., M.S., and R.R. 
acknowledge support by the Deutsche Forschungsgemeinschaft (DFG) within 
the FOR~2440. B.B.L.W., M.S., and S.H. were supported by the DOE Office of Science, Fusion Energy Science under FWP~100182.
The work of T.D. and P.A.S. was performed under the auspices of the U.S. Department of Energy by Lawrence 
Livermore National Laboratory under Contract No. DE-AC52-07NA27344 and supported by Laboratory Directed Research 
and Development (LDRD) Grant No. 18-ERD-033. D.K. was supported by the Helmholtz Association under VH-NG-1141.
The DFT-MD calculations were performed at the North-German Supercomputing Alliance (HLRN) facilities and the 
computing cluster Titan hosted at the ITMZ at University of Rostock.

\bibliography{ionization.bib}

\end{document}